\documentclass[12pt]{article}
\usepackage{hyperref}
\begin{document}

\title{Towards A Grand Unification}
\author{Burra. G. Sidharth\footnote{gsburra7748@gmail.com, University of 
Udine, Italy}}

\date{}
\maketitle

\begin{abstract}
The author had shown that gravitation can be reconciled with
electromagnetic and other forces if we start from a Landau-Ginzburg
phase transition. This is further remarked upon and a unification of all forces 
of nature is proposed. 
\end{abstract}
\maketitle

\section{Introduction}

The notion of a unified theory that is comprised of all the fundamental forces 
of nature has it's roots since the time of Einstein,
Kaluza, Hermann Weyl and others. Several eminent physicists have proposed 
various methodologies to model a
unified interaction that is given by a single gauge symmetry of several force 
carriers. There is a vast amount of literature 
\cite{Pati,Georgi,Ellis,Gouvea,Aguila} (and several references therein) on this 
topic of widespread interest. \\
Our approach is distinct from the various other approaches  used.  However, we 
would like to point
out that this need not be so.\\
Way back in 2003 the author had pointed out that if we start with a
Landau-Ginzburg phase transition, then we go from a Planck scale
of gravitation to the Compton scale of other interactions
\cite{csf2003,epj,tduniv}. It is the purpose of this paper to show that one may 
unify all the four forces of nature. We start with  a unified Lagrangian and 
it's characteristics corresponding to  phase transition. It has been shown 
explicitly that gravitation and electromagnetism are intrinsically connected. 
\section{The Unified Lagrangian}

Although, the Ginzburg-Landau theory \cite{Land} was a phenomenological model 
used to describe superconductivity it has a far-reaching influence in 
understanding some fundamental microphysical phenomena. We commence by 
considering the Landau-Ginzburg Lagrangian that is
given by
\begin{equation}
{\bf L}_{GL} = L_{0} + a|\psi|^{2} + \frac{b}{2}|\psi|^{4} +
\frac{1}{2m}|(\frac{\hbar}{i}\nabla - e{\bf A})\psi|^{2} + \frac{{\bf
B}^{2}}{2\mu_{0}}\label{A}
\end{equation}
It has been discussed in detail that the Lagrangian (\ref{A}) via a
phase transition leads to the Compton scale of other interactions,
starting from the Planck scale which describes gravitation.\\
Let us now write a grand unified Lagrangian given by
\begin{equation}
{\bf L}_{U} = \alpha(t) {\bf L}_{GL} + \beta(t) {\bf
L}_{SM}\label{B}
\end{equation}
where $\alpha(t)$ and $\beta(t)$ are time dependent coefficients and ${\bf
L}_{SM}$ is the standard model Lagrangian given as\\
\[{\bf L}_{SM} = {\bf L}_{Dirac} + {\bf L}_{Mass} + {\bf L}_{gauge} + {\bf L}_{gauge/\psi}\]\\
Now it is known that the standard model Lagrangian in (\ref{B})
leads to the electro weak and strong interactions.\\
Thus a single Lagrangian ${\bf L}_{U}$ given in (\ref{B}) describes
all four interactions viz., the gravitational and other three
interactions. This is the desired result.\\
Now, the unified Lagrangian (\ref{B}) can be looked upon as a superposition of 
the Lagrangians ${\bf L}_{GL}$ and ${\bf L}_{SM}$, in analogy with the concept 
of `superposition of states' in quantum mechanics. One thing should be borne in 
the mind that the coefficients $\alpha(t)$ and $\beta(t)$ cannot be zero at the 
same epoch '$t$', for that would make the Lagrangian ${\bf L}_{U}$ trivial. 
But, both of the coefficients are allowed to take the value zero.\\
In some epoch $t_{1}$, suppose $\alpha(t_{1}) = 0$ and $\beta(t_{1}) \neq 0$ then we have\\
\[{\bf L}_{U} = \beta(t) {\bf L}_{SM}\]\\
comprised of the symmetry $SU(3) \times SU(2) \times U(1)$. Thus in the epoch 
$t = t_{1}$ the unified Lagrangian ${\bf L}_{U}$ has only the standard model. 
Again, in some epoch $t_{2}$ if we suppose that $\alpha(t_{2}) \neq 0$ and 
$\beta(t_{2}) = 0$ then we have\\
\[{\bf L}_{U} = \alpha(t) {\bf L}_{GL}\]\\
which gives us only gravitation. Thus, the unified Lagrangian ${\bf L}_{U}$ has 
only gravitation. This justifies why the Lagrangian (\ref{B}) gives a unified 
description of all the four known interactions.\\
Now, it is interesting to note another aspect of the Lagrangian (\ref{B}). By 
virtue of the universality of critical phenomena we know that gauge theories 
can demonstrate different phases \cite{Paul}. Quantum electrodynamics, weak 
force, strong force all exhibit different phases. Now, it was shown by the 
author  that at the Planck scale we have only gravitation and at the Compton 
scale all other forces emerge. This does not mean that at the Compton scale 
gravitation vanishes. Actually, at the Compton scale gravitation becomes very 
weak. 
It is also conspicuous because of the fact that the {\it Planck length} 
$\approx 10^{-33} cms$ and the {\it Compton length} $\approx 10^{-11} cms$. 
This also explains why the electromagnetic force is many times stronger than 
the gravitational force. This aspect will also be discussed in the next 
section.\\
If measurements are made at the Planck scale for the Ginzburg-Landau Lagrangian 
then one can get back gravitation very easily. This was also proved by the 
author. It is obvious that the length scales are extremely fundamenal in 
nature. Here, one could use to the concept of the {\it renormalization 
group} \cite{Wilson1,wil} and the universal nature of critical phenomena. It is 
justified if we say that while the transition from the Planck scale to the 
Compton scale there is a critical point where the various physical parameters 
reach their critical value. Therefore, one must use the renormalization group 
methodology so that the effects of the length scale below the {\it correlation 
length} $\xi$ \cite{Wilson1} gets averaged out and we get the proper physical 
results in the Compton scale. \\
In executing this task one can achieve a fixed point or equilibrium point where 
the unified Lagrangian (\ref{B}) becomes stable and all the forces of nature 
are unified. Thus, the technique is a good tool for the unification of all the 
forces even though they might be in different phases. In fact, it is easy to 
see that using the exact method of Wilson \cite{Wilson1,Wilson2}, one can 
arrive at a modified Ginzburg-Landau Lagrangian.\\
Now, we would like to follow  Wilson and considering universality of critical 
phenomena we rewrite the Ginzburg-Landau Lagrangian (\ref{A}) simply as\\
\[{\bf L} = V[a\psi^{2}(x) + \frac{b}{2}\psi^{4}(x)]\]\\
Here, the magnetization has been replaced by the wavefunction in contrast to 
Wilson's work and $V$ is an arbirary volume of space where we  carry out our 
task. We presume that the coefficients $a$ and $b$ are dependent on a 
fundamental cut-off length ($l$)  and rewrite them as $a_{l}$ and $b_{l}$ 
respectively. Similarly, we rewrite the Lagrangian density as\\
\begin{equation}
{\bf L}_{l} = V[a_{l}\psi^{2}(x) + \frac{b_{l}}{2}\psi^{4}(x)] \label{d}
\end{equation}
Now, in analogy to the case of magnetization, we presume that there would remain effects of fluctuations in $\psi(x)$ even when measurements are done above the cutoff length. More precisely, in the infinitesimal interval $l + \delta l$ the effects of the lengths between $l$ and $l + \delta l$ will persist to linger on. This is where we average out such effects in analogy with Wilson \cite{Wilson1}. The immediate consequence of such averaged out fluctuations would modify the Lagrangian density as $L_{l+\delta l}$, where the modified wavefunction would be $\psi_{m}(x)$. \\
Again, considering localized wavepackets in accordance with Wilson, we may write\\
\begin{equation}
\psi(x) = \psi_{m}(x) + \sum_{n}r_{n}\phi_{n}(x) \label{e}
\end{equation}

where, each wavefunction $\phi_{n}(x)$ fills a unit volume in the phase space. 
Here, the integration is performed with respect to  the coefficients $r_{n}$. 
Now, considering a single integration we write\\
\begin{equation}
\exp[-{\bf L}_{l+\delta l}(\psi_{m})] = \int_{-\infty}^{+\infty} \exp[-{\bf L}_{l}(\psi_{m} + r\phi)] {\rm d}r
\end{equation}
where, ${\bf L}_{l+\delta l}$ and ${\bf L}_{l}$ involve integration only over the volume occupied by the function $\phi(x)$. We presume that $\psi_{m}(x)$ is independent of the coefficients $r$ and thus we have\\
\begin{equation}
\exp[-{\bf L}_{l+\delta l}(\psi_{m})] = \exp[-{\bf L}_{l}] \int_{-\infty}^{+\infty} \exp[g(l)r^{2} + h(l)\psi^{2}r^{2}] {\rm d}r
\end{equation}
where, $g(l)$ and $h(l)$ are functions of the length $l$ and consist of the coefficients $a_{l}$ and $b_{l}$ of the Lagrangian density (\ref{d}). Thus we may write \\
\begin{equation}
{\bf L}_{l+\delta l}[\psi_{m}(x)] = {\bf L}_{l}[\psi_{m}(x)] + \frac{1}{2}\ln[g(l) + h(l)\psi_{m}^{2}(x)]
\end{equation}
Therefore, for the $l + \delta l$ range of lengths and one integration, the Ginzburg-landau Lagrangian density from (\ref{d}) can be written as\\
\begin{equation}
{\bf L}_{l+\delta l}[\psi_{m}(x)] = a_{l}\{\psi_{m}(x) + r\phi(x)\}^{2} + \frac{b_{l}}{2}\{\psi_{m}(x) + r\phi(x)\}^{4} + \frac{1}{2}\ln[g(l) + h(l)\psi_{m}^{2}(x)] \label{f}
\end{equation}
This can be extended to the entire volume space $V$, where the contributions of 
all the coefficients $r_{n}$ are included. Now, it is obvious from this result 
that the Ginzburg-Landau Lagrangian density in the unified Lagrangian gets 
slightly modified owing to the effects of the fluctuations arising due to 
contributions from the lengths between $l$ and $l + \delta l$, and their 
averaging. Similarly, the standard model Lagrangian must be modified in order 
to get precise results. In fact, one may consider the length $l$ to be the 
Compton length $l_{c}$ and the various problems in unifying all theories might 
be truncated.\\
Another thing that should be borne in the mind is that since the Ginzburg-Landau Lagrangian gets modified due to the averaging effects it is obvious that the equations of motion and the related dynamics will get modified. Consequently, the energy spectrum of states and other such fundamental aspects might need a new viewpoint.\\
Now, since we are considering fluctuations and their effects we may consider Ito's lemma \cite{Ito1,Ito2} as\\
\[{\rm d}\psi(x_{t}) = \psi^{\prime}(x_{t}){\rm d}x_{t} + \frac{1}{2}\psi^{\prime\prime}(x_{t})\sigma^{2}_{t}{\rm d}t\]\\
where, the {\it primes} denote derivative with respect to $x_{t}$, $\sigma_{t}$ is the standard deviation and the subscript refers to the instant of time. From this we can obtain a covariant derivative of the form\\
\begin{equation}
{\rm d}\psi_{m}(x_{t}) = \{\psi^{\prime}_{m}(x_{t}) + \sum_{n}r_{n}\phi^{\prime}_{n}(x_{t})\}{\rm d}x_{t} + \frac{1}{2}\{\psi^{\prime\prime}_{m}(x_{t}) + \sum_{n}r_{n}\phi^{\prime\prime}_{n}(x_{t})\}\sigma^{2}_{t}{\rm d}t - \sum_{n}{\rm d}\{r_{n}\phi_{n}(x_{t})\} \label{g}
\end{equation}
This covariant derivative might be able to describe the interaction arising after the phase transition, namely electromagnetism. Interestingly, the sudden change in the scale, i.e. from the Planck scale to the Compton scale can be looked upon as a {\it Heavyside step function} like behaviour. Specifically, the function $\phi(x)$ in (\ref{e}) can be thought of as a similar step function such that\\
\begin{equation}
\phi_{n}(x) =
     \cases{0, & $n < 0$\cr\\
     1, & $n \geq 0$.}
\end{equation}
Here, we would like to delineate a few aspects of this {\it step function like} behaviour. Here, \\
1)~$n < 0$ means that there are no modes of fluctuations.\\
2)~$n = 0$ means that it represents the zeroth mode, i.e. $\phi_{0}(x)$.\\
3)~$n > 0$ means that the number of modes of fluctuations are greater than 0.\\
4)~$\phi_{n}(x) = 0$ means that there are no effects fluctuations, i.e. when we are still in the microscopic scale (in the $l$ to $l + \delta l$ region).\\
5)~$\phi_{n}(x) = 1$ means that effects of fluctuations are present, i.e. when we are in the comparatively macroscopic scale (above the $l + \delta l$ region).\\
Now, we know that the derivative of the {\it Heavyside step function} can be defined as the {\it Dirac delta function}. Considering the length scale we write\\
\[\frac{{\rm d}\phi(x)}{{\rm d}l} = \delta(l)\]\\
Now, keeping in mind equation (\ref{g}) we write this as\\
\begin{equation}
\frac{{\rm d}\phi(x_{t})}{{\rm d}x_{t}}\frac{{\rm d}\phi(x_{t})}{{\rm d}l} = \delta(l)
\end{equation}
If {\it primes} denote derivatives with respect to $x_{t}$ then we have\\
\begin{equation}
\phi^{\prime}(x_{t}){\rm d}x_{t} = \delta(l){\rm d}l \label{h}
\end{equation}
This means that the terms $(r_{n}\phi^{\prime}_{n}(x_{t}){\rm d}x_{t})$ and $(r_{n}\phi^{\prime\prime}_{n}(x_{t})\sigma^{2}_{t}{\rm d}t)$ of (\ref{g}) will have a delta function type contribution in the Lagrangian (\ref{f}) when it is considered over the whole volume $V$. Interestingly, we know that the delta function has the heuristic characterization\\
\begin{equation}
\delta(l) =
     \cases{\infty, & $l = 0$\cr\\
     0, & $l \neq 0$.}
\end{equation}
Now, suppose we take the Planck scale into consideration. Since, the Planck length ($10^{-33}cms$) is very small we might take it to be almost zero. So, when the corresponding length ($l$) $\approx 0$, we have a large delta function type contribution. Albeit, when (\ref{g}) is integrated we will be left with the contribution of the coefficients $r_{n}$ in the Lagrangian (\ref{f}) because of the property of the delta function\\
\[\int_{-\infty}^{+\infty}\delta(l){\rm d}l = 1\]\\
On the contrary, we shall see that the terms $(r_{n}\phi^{\prime\prime}_{n}(x_{t})\sigma^{2}_{t}{\rm d}t)$ of (\ref{g}) is much more interesting. Now, equation (\ref{h}) can also be written as\\
\[\phi^{\prime}(x_{t}) = \delta(l)\frac{{\rm d}l}{{\rm d}x_{t}}\]\\
Therefore, differentiating this relation with respect to $x_{t}$ we have\\
\begin{equation}
\phi^{\prime\prime}(x_{t}) =  \delta^{\prime}(l)\frac{{\rm d}l}{{\rm d}x_{t}} + \delta(l)\frac{{\rm d^{2}}l}{{\rm d}x_{t}^{2}} \label{i}
\end{equation}
Again, we know that in electromagnetism \cite{Jackson,Zangwill}, the
gradient of the delta function represents a point magnetic dipole
situated at the origin and that the function itself represents a
point charge \cite{Glen}, owing to it's distributional property.
Thus, we are in a position to infer that the two terms of (\ref{i})
comprised of the first derivative of the delta function and the
function itself give rise to the phenomenon of electromagnetism.
This, in a rigourous way substantiates the author
Sidharth's earlier work \cite{csf2003} which stated that via a Ginzburg-Landau phase transition one can easily go from graviatation to electromagnetism. Of course, everything boils down to the length scale under consideration. Ostensibly, this particular term signifies the sharp contrast between the gravitational force and the other three forces.\\
Equation (\ref{e}) and equation (\ref{g}) are crucial for this
perception. Without the renormalization group technique and the
fluctuations of the lower length scales that give rise to (\ref{g})
we would not have achieved this.
Thus, it is prerequisite that one takes into consideration these two factors when considering any grand unification scheme.\\
To be more subtle, we may infer that the Ginzburg-Landau Lagrangian represents a symmetry since we have used the renormalization group scheme to find the connection between gravitation and electromagnetism. This Lagrangian acts as the bridge between gravitation and the other three forces of nature.

\section{The Fundamental Length Scaling Factor ({\it FLSF})}

Now, from the preceding section we have the grand unified Lagrangian given as
\begin{equation}
{\bf L}_{U} = \alpha(t) {\bf L}_{GL} + \beta(t) {\bf
L}_{SM}
\end{equation}
Here, suppose we choose the coefficient $\alpha(t)$ as\\
\begin{equation}
\alpha(t) = \ln[f_{GL}(t)] \label{C}
\end{equation}
where, $f_{GL}(t)$ is another function dependent on time. We define it as\\
\begin{equation}
l_{t} = f_{GL}(t)l_{0}
\end{equation}
Here, $l_{t}$ is the fundamental minimum length at an epoch '$t$' and $l_{0}$ is the fundamental minimum length at a reference time '$t_{0}$'. The function $f_{GL}(t)$ is defined as the {\it fundamental length scaling factor} (FLSF) for the Ginzburg-Landau Lagrangian. This factor scales the fundamental length scale at different epochs. Now, from the above equation we have at $t = t_{0}$\\
\[f_{GL}(t) = 1\]\\
and consequently\\
\[\alpha(t) = 0\]\\
Thus, at a reference time $t = t_{0}$ we have the Lagrangian as\\
\begin{equation}
{\bf L}_{U} = \beta(t) {\bf
L}_{SM}
\end{equation}

which is nothing but the standard  model Lagrangian. The scaling factor (FLSF) is interesting in the sense that it determines the fundamental minimum length related to the Ginzburg-Landau Lagrangian by scaling it at different epochs. As for example, suppose at a certain epoch '$t_{1}$' the fundamental length was the {\it Planck length}\\
\[l = l_{p} \approx 10^{-33} cms\]\\
and at the current epoch '$t_{2}$' it is the {\it Compton length}\\
\[l = l_{c} \approx 10^{-11} cms\]\\
Thus, for the two aforesaid cases we have the FLSF respectively as\\
\begin{equation}
f_{GL}(t_{1}) = \frac{l_{p}}{l_{0}}
\end{equation}
and
\begin{equation}
f_{GL}(t_{2}) = \frac{l_{c}}{l_{0}}
\end{equation}

Comparing these two equations we have\\
\begin{equation}
\frac{f_{GL}(t_{1})}{f_{GL}(t_{2})} = \frac{l_{p}}{l_{c}}
\end{equation}
Therefore, the ratio of the FLSF for the two epochs '$t_{1}$' and '$t_{2}$' is given as\\
\begin{equation}
\frac{f_{GL}(t_{1})}{f_{GL}(t_{2})} = 10^{-22}
\end{equation}
In this manner, one can compare the scaling factors in different epochs, or from a known ratio value one may be able to determine the ratio of the fundamental minimum lengths. Now, as we have argued before the different length scales contribute to the difference in the various interactions, specifically gravitation and the other three forces. Such difference between the {\it Planck scale} and the {\it Compton scale} supports the well known fact, as to why there is a monumental difference between the gravitational force and the other three forces. Everything comes down to the scale we are considering to explain the various physical phenomena and the fact that there is a Ginzburg-Landau phase transition at the heart of this major difference.\\

Now, by the same method we can arrive at the FLSF for the standard model Lagrangian. Only, since the coefficients $\alpha(t)$ and $\beta(t)$ cannot be zero at the same epoch $t$ we choose instead of (\ref{C})\\
\begin{equation}
\beta(t) = \ln[f_{SM}(t) + a]
\end{equation}
where, $a$ is some constant independent of time and $f_{SM}(t)$ is the FLSF for the standard model Lagrangian that scales the corresponding fundamental minimum length at various epochs. Nonetheless, one must remember that the fundamental minimum length scale is just a limit. Although, in an arbitrary epoch there might be a fundamental length, we must consider the effects of all lengths below the fundamental length. This is the key essence of this paper and the key concept of the renormalization group methodology by Wilson.\\
It must be borne in the mind that neglecting the fluctuational effects and not scaling them would culminate in wrong perception of the physical world. This is what has deterred scientists from producing a unified theory of all forces. But, our efforts ingrained in this paper might be able to prove otherwise.

\section{Discussions}

In the present paper we have proposed a new scheme of a unified theory by using 
Wilson's renormalization group methodology. The Ginzburg-Landau Lagrangian 
introduced at the beginning of the paper can be looked upon as the 
representative of a new symmetry that connects gravitation and the three gauge 
interactions. The exact Lie algebra, the generators, the operators and their 
properties corresponding to this renormalization group technique is beyond the 
scope of the current paper. But is a good topic for future research.\\
The importance of the {\it Fundamental Length Scaling Factor} (analogous to the 
cosmic scaling factor), in terms of the epoch of the universe has been 
investigated in a quantitative manner. A detailed qualitative analysis of this 
parameter might prove as a key ingredient to conceive the concept of 
unification.\\

As already mentioned, there is a bispinorial underpinning for space time 
\cite{BGS-0106051}. This arises because, as explained in the reference, there 
are actually two Weyl  spinors which have opposite behaviour under reflection. 
So, near the Compton wavelength, we have the bispinorial space. At distances 
far from the Compton wavelength we return to the usual Minkowski spacetime. So 
gravitation and electromagnetism get a joint description and further if we use 
the electro weak unification –– we have here a \textit{theory of everything.}

\end{document}